# Flood avalanches in a semiarid basin with a dense reservoir network

Samuel J. Peter, J. C. de Araújo, N. A. M. Araújo, H. J. Herrmann

**Abstract**. This study investigates flood avalanches in a dense reservoir network in the semiarid north-eastern Brazil. The population living in this area strongly depends on the availability of the water from this network. Water is stored during intense wet-season rainfall events and evaporates from the reservoir surface during the dry season. These seasonal changes are the driving forces behind the water dynamics in the network. The reservoir network and its connectivity properties during flood avalanches are investigated with a model called ResNetM, which simulates each reservoir explicitly. It runs on the basis of daily calculated water balances for each reservoir. A spilling reservoir contributes with water to the reservoir downstream, which can trigger avalanches affecting, in some cases, large fractions of the network. The main focus is on the study of the relation between the total amount of water stored and the largest observable cluster of connected reservoirs that overspill in the same day. It is shown that the thousands of small and middle-sized reservoirs are eminent for the retention of water upstream the large ones. Therefore, they prevent large clusters at a low level of water. Concerning connectivity measures, the actual reservoir network, which evolved without an integrated plan, performed better (i.e., generated smaller avalanches for similar amount of stored water) than numerous stochastically generated artificial reservoir networks on the same river network.

**Key-words:** scale-free network, flood propagation, connectivity

## 1. INTRODUCTION

The north-east of Brazil is a semiarid region characterized by intermittent precipitation events during rainy season and long periods of water scarcity. Reoccurring droughts, sometimes consecutively for several years, produce serious socioeconomic damage. The water – mainly needed for irrigation, human, and animal uses – is vital for the farmers and its availability is at risk during dry periods (Gaiser et al., 2003). To overcome the problem of the low availability of natural water and the high water demand due to dense population, authorities, land owners, and communities have built thousands of on-river water reservoirs of a wide range of sizes, which allow the water stored during the rainy periods to be used during the dry season. Most of the reservoirs have been constructed according to the local needs of the population, especially farmers, without an integrated plan. This has resulted in a complex dense reservoir network, which is extremely difficult to manage (de Araújo et al., 2010; Krol et al., 2011; Lima Neto et al., 2011; Malveira et al., 2012).

Investigations of the hydrological impact of the small reservoirs on large scale water availability have been conducted (Krol et al., 2011; Malveira et al., 2012) using the WASA model (Water Availability in Semi-Arid Environments see, for example, Güntner and Bronstert, 2004; Güntner et al., 2004; Krol et al., 2006) in Brazil and the TEDI simulation model (Tool for Estimating Dam Impacts see, for example, Nathan et al., 2005; Lowe et al., 2005) in Australia. WASA has been developed as a deterministic, spatially distributed model, to quantify water availability in large water-scarce regions. In WASA, the small and middle-sized reservoirs are represented in classes and, therefore, the water balance of these reservoirs is not calculated explicitly. As a first approach, the hydrologic impact of small dams might be quantified as the sum of the single reservoirs. However, the properties of large-scale system cannot be anticipated due to the emergent behavior arising from the interaction between single reservoirs, as Mamede et al. (2012) showed in the case of the overspill avalanches. In that work, self-organized criticality (SOC) was found within the reservoir network. The concept of SOC states that complex behavior can develop in multi-body systems without explicit pressure or constraints from outside the system, evolving either in time or space (Bak, 1996). In the present, we go beyond to study the effect of the network topology on the overall dynamics. In particular, we test different network configurations and analyze the water dynamics within these topologies.

To take reasonable decisions within the framework of water resources management at the level of the reservoir network, instead of looking at single reservoirs, an integrated analysis of the entire network is needed. The motivation for this study is to understand the dependence of water availability on the connectivity pattern, since the two components are correlated: The more water is available, the higher is the connectivity among the reservoirs, and thus the more vulnerable is the network to floods. The concept of connectivity in hydrological systems had been widely applied and it is seen as the key driver for many catchment processes (Michaelides and Chappell, 2009). In the present study, connectivity is defined as the efficiency with which water is transported downstream through the network, which is highly influenced by the thousands of reservoirs in the network. As none of the reservoirs features gates to operate in case of flooding, the natural effect of the reservoirs, which reduces the peak discharge of floods downstream, is the only tangible effect of the network on the flood routing. Therefore a better understanding of the role of connectivity is expected to help mitigating flooding problems, improve water quality, and address economic questions. This might be achieved by changing the system when adding or removing specific reservoirs, or changing the attributes.

By applying the method of upscaling, by which the information of small scale units is used to answer questions at a larger scale, a new model has been developed, denoted ResNetM. Following the approach of network theory, the model provides a simple water balance for each reservoir, as well as information about the water distribution in the system over decades (Mamede et al., 2012). A spilling reservoir delivers water to its downstream reservoir, which can lead to further spilling in a cascade way. These spilling connections among reservoirs can lead to catastrophic events extending over large fractions of the reservoir network (Mamede et al., 2012), with severe socioeconomic consequences (Dutta et al., 2003; Zhou et al., 2012).). It is, therefore, paramount to understand the dynamics with respect to these spilling avalanches.

The area under focus is the Upper Jaguaribe basin, located in the south of the state of Ceará, Brazil, which is part of the Brazilian semiarid drought polygon (see Figure 1). The Upper Jaguaribe consists of 24 municipalities with approximately half a million inhabitants. The economy is mainly built on cattle breeding and agriculture, both irrigated and rain-fed. Its catchment area extends over more than 24000 km² and contains between 4000 and 5000 reservoirs distributed over the basin. The outlet of the basin is controlled by its largest reservoir, Orós, which was built in 1961 and has a storage capacity of 1.94 billion m³. The average annual rainfall in the region is around 780 mm, whereas annual class A pan evaporation rate is three times as high (2500 mm). Despite the numerous reservoirs, a maximum of 150 mm, or 20% of annual precipitation, has been stored in the system. The groundwater resources are poor and often salty because of the crystalline bedrock (Andrade et al., 2008). In the south and in the east of the basin, small regions with sedimentary rocks can be found. The shallow soils, the reduced vegetation cover, and the high temporal variability in rainfall lead to ephemeral or intermittent rivers (Gaiser et al., 2003). The coefficients of variation are up to 1.4 for annual river discharge. Typical runoff coefficients are in the range of 7%, but values as low as 3% have been identified, depending on geological constraints (de Araújo and Piedra, 2009).

Among the thousands of dams there are a few large ones referred to as *strategic*, because they are planned to secure water supply even in the case of several consecutive drought years. Well-designed reservoirs typically yield 40% of the annual direct inflow with an annual reliability of 90% (Campos, 2010; de Araújo et al., 2006). On the other hand, the smallest ones dry out even before the end of the dry season. On average, the reservoirs are between 50 and 70% full at the end of the rainy season and reach a level down to approximately 30 – 40% of their capacity at the end of the dry season. Although the thousands of small reservoirs are hydrologically less efficient due to their open morphology, they are very important for a democratic distribution of the water (Malveira et al., 2012). They are also energetically efficient, because they raise the water gravity center, demanding less energy to pump water to upstream consumers. In addition, they show a positive impact on the sedimentation of the strategic reservoirs by detaining large amount of sediments further upstream (Lima Neto et al., 2011).

## 2. MATERIAL AND METHODS

After an outline of the model and data that was used, the main indicators to quantify the systems behavior are then introduced. The subsequent section defines additional methods to evaluate the performance of the reservoir network in the context of water availability and flood-avalanche management. In the end the model setups that are subject to the described evaluations are delineated in detail.

### 2.1. The ResNetM model

Malveira et al. (2012) and Krol et al. (2011) have used the WASA model to analyze the Upper Jaguaribe reservoir network. Despite its robust physical base, the WASA model considers the small reservoirs non-explicitly, i.e., the several thousands of small and middle-sized reservoirs are represented in an aggregated way by grouping them into five classes depending on their storage capacity (Güntner et al., 2004). Contrastingly, the model used hereafter (ResNetM) can be characterized by simplicity in representing the physical processes, but complexity in representing the network topology, i.e., each and every reservoir and river reach are explicitly considered and geo-referenced. The main physical mechanisms considered are the runoff generation and the evaporation from the reservoir free surface. This way, few watershed parameters are needed, whereas a considerable amount of topological parameters is required. The resulting computation time is low, which opens the possibility for intensive studies of the parameter space, of the network spatial distribution and of reservoirs interplay. The water balance at each reservoir is given by

$$V_{i,t+\Delta t} = V_{i,t} + H_{i,t} \cdot (A_{i,t} \cdot RC_{i,t} + S_{i,t}) + \sum_{j=1}^{N_i}[Qu_{i,j,t} \cdot (1 - Kt_{j,i}) \cdot \Delta t] - E_t \cdot (Kev + Kwu) \cdot S_{i,t+\Delta t/2} - Q_{i,t} \cdot \Delta t, \quad (1)$$

where $V_{i,t}$ is the volume at the present reservoir *i* at the present time step *t*; *H* is the rainfall; *A* is the respective direct catchment area; *RC* is the runoff coefficient; *S* is the reservoir water surface; *Qu* is the discharge contribution of directly upstream reservoirs (there are $N_i$ reservoirs directly connected with reservoir *i*); *Kt* is the non-dimensional parameter for transmission losses (between zero and one); *E* is the class A pan evaporation rate; *Kev* the coefficient that represents the ratio between evaporation in the lake and in the pan; *Kwu* is the coefficient of water use; and *Q* is the spilling discharge of the reservoir. To calculate the amount of evaporated water, the water surface of the reservoir is taken as an average between the surface at the beginning of the time step and the surface after taking into account all inflows (upstream reservoirs and rainfall) and is therefore denoted as $S_{i,t+\Delta t/2}$ in (1).

At each time step *Δt* the water balance is calculated for every reservoir, integrating from upstream to downstream. Since the routing time between the reservoirs is not considered in the model, floods always propagate from the network boundaries downstream to the outlet of the system within one time step. Therefore the choice of *Δt* has no influence on the emergence of flood avalanches. This is reasonable because temporal aspects of the system are not of interest in the context of this investigation. The integration time step chosen here is one day, since the data available for rainfall is at a resolution of one day as well. Because the volume of each reservoir at the new time step is only dependent on variables of the previous time step, the routine is explicit and numerically simple to implement. The simulation time for twenty years and a network of about 5'000 reservoirs is around one minute. The model is expected to capture both short time events (flood avalanches) as well as long time characteristics (water being stored) of the reservoir system.

The lake evaporation rate is assumed to be proportional to the measured class A pan rate (*Kev* is the constant of proportionality), whose typical values in semiarid regions are around 0.7 (Gundekar et al., 2008). The water uses, such as irrigation and residential water supply, are assumed to be proportional to the evaporation rate as well. More specifically they are considered proportional to class A pan evaporation, where *Kwu* ~ 0.35, as calculated for several semiarid reservoirs (de Araújo et al., 2006). For the sake of simplicity, we admitted *Kev* + *Kwu* ~ 1.0 in ResNetM.

Important hydrological processes, such as interception and watershed infiltration, are implicitly captured in the concept of the runoff coefficient (Maidment, 1993). When applying ResNetM, the runoff coefficient can be either kept constant over time and space, or can be a function of the recent history of rainfall. Rainfall leads to an increase in the soil moisture, reducing the potential infiltration rate and, consequently, enhancing the runoff coefficient. Besides, consecutive events might result in the formation of base-flow, which – otherwise – is either negligible or null in the intermittent-river system of the Brazilian semiarid region (Medeiros et al., 2010; de Araújo and Piedra, 2009). The time evolution of the runoff coefficient RC is given by

$$RC_{i,t+\Delta t} = RC_{i,t} \cdot K_{RC}, \tag{2}$$

where $K_{RC}$ is a coefficient that represents the temporal changes of *RC* depending on the soil moisture. For rainy days $K_{RC} > 1$; whereas the runoff coefficient decreases ($K_{RC} < 1$) for dry days.

A reservoir spills whenever its storage capacity is exceeded. It is crucial to adequately compute the surface – volume relation *S(V)* of the reservoirs, among others, to estimate the evaporation flux. A study of 417 reservoirs in the Brazilian semiarid region (Molle, 1989; Molle and Cadier, 1992) proposes a relation given by

$$S = c \cdot d \cdot \left(\frac{V}{d}\right)^{\frac{c-1}{c}}, \tag{3}$$

where *c* is a parameter describing the shape of the valley in which the reservoir is located, and *d* stands for the coefficient dependent on the angle of aperture of the valley. The parameters *c* = 2.7 and *d* = 1500 were empirically established and tested for hundreds of small dams of different construction ages; and proved valid for the ensemble (Molle, 1989). The parameter *c* corresponds to a V-shaped valley and a regular angle of aperture for reservoirs used for irrigation purpose. Thus, the surface scales like *S* ~ $V^{0.63}$. Equation (3) showed good agreement with measured data of 21 reservoirs of the Upper Jaguaribe basin (Mamede et al., 2012: data provided by COGERH, 2011).

When spilling occurs, the transmission losses in semiarid intermittent rivers can be eminent in hydrological modeling (Costa et al., 2012; Boroto and Gôrgens, 2003; Pilgrim et al., 1988). In ResNetM this is implemented in a purely conceptual way, where a dual parameter describing the transmission losses is depending on whether the river reach between the upstream and the downstream reservoirs was dry or wet in the previous day.

**2.2. Hydrological data**

The rainfall data is available from the website of the governmental organization FUNCEME (2011). For the simulations in the area of the Upper Jaguaribe basin, the time series of rainfall from 131 gauging sites is considered. The use of as many spatially distributed rainfall gauges as possible is justified by the high rainfall-intensity variability in semiarid regions (García-Pintado et al., 2009). In the case of annual rainfall, the temporal variation is rather high, and the coefficient of variation amounts to 30%. For example, in the year of 1985 there was more than four times as much rain as two years before. The intra-annual variability is extremely high, with clear distinction between the wet and dry seasons. Almost 80% of annual rainfall occurs in the period from January to April, which will be hereafter referred to as wet season; whereas the remaining months form the dry season. The evaporation rate is taken only from one climate station, Campos Sales, because the spatial fluctuations are very low. Over the year the rate has its peak antipodal to the annual rainfall cycle during the months of August and September.

**2.3. River and reservoir networks**

2.3.1. River network

The reservoirs are connected through the drainage network which, in this research, was obtained from the analysis of high resolution elevation data, collected by the Shuttle Radar Topography Mission (SRTM, 2011) from NASA. The data is represented by a rectangular grid with a resolution of 90 m × 90 m. The delineation of the drainage network is accomplished using the method described by Jenson and Domingue (1998; see also O'Callaghan and Mark, 1984): for each cell a single flow direction is assigned in the direction of the steepest local slope (see, e.g., model Dicasm: Montenegro and Ragab, 2012). The extraction of the river network from the drainage network is essential to detect the true reservoirs. The channel initiation method with a constant threshold value was applied (Tarboton et al., 1991), for which the threshold area for the channel head is fixed at ten cells (0.081 km²). A similar value has been chosen for the same watershed in a previous investigation (Mamede et al., 2012) and a visual comparison with maps showed good agreement.

2.3.2. Reservoir network and outliers elimination

The identification of the reservoirs, together with their maximal water surface and geo-referenced location, were obtained from satellite pictures imagery immediately after the rainy season for three very wet years, namely, 2004, 2008, and 2009. It was assumed that all reservoirs were at their maximum capacity after these intensive rainfalls. Possible sources of error in the identification process are clouds on the satellite pictures and the fact that it is not sure that all reservoirs were at their maximal water level when the satellite pictures were taken. In addition, it is usual that large and flat areas get flooded during heavy rainfall events, which can be misinterpreted as reservoirs and get a capacity assigned larger than the actual volume of the flooded area. The data concerning the 19 monitored reservoirs – whose individual storage capacity ranges from 2 to almost 2 000 hm³, and whose global capacity adds to 2810 hm³ – were obtained in official files (COGERH, 2011) and used in this research.

The originally collected data consists of 4750 reservoirs (see Table 1 and Figure 9). The outlet of each reservoir was determined by the pixel having the largest drainage area. In most cases, the outlet was located directly on the extracted river network. If not, then the access to the river network could be found by following the drainage route. The directly downstream connected reservoir was assigned by routing through the river network. Starting from the system with all reservoirs (network A), five other networks were generated by appropriately removing outliers manually (networks B and C); or automatically (networks D, E and F), as shown in Table 1. Outliers stand for reservoirs that are supposed to be rather wide river streams or flood plains than actual reservoirs. These modifications were made due to the uncertainties in the data assessment. In networks B and C, the outliers were visually identified by two different experts, as in Malveira et al. (2012). In network F, all reservoirs with a distance between outlet and river network larger than two grid cells were dismissed. In addition, reservoirs having a ratio between capacity and total drainage area smaller than 1 mm were deleted in network E; and additional reservoirs in network D were eliminated, which had a ratio between capacity and direct drainage area (i.e., disregarded the drainage area of the upstream reservoirs) smaller than 1 mm.

## 2.4. Quantifying the connectivity of the reservoir system

We define avalanche size ($x$) as the number of connected reservoirs that overspill in the time scale of one day, i.e. after each calculation step every obtained cluster of at least two overspilling connected reservoirs is listed in a frequency table. Given a certain time horizon ($T$), the normalized list of accumulated reservoir clusters is defined as the avalanche size distribution. As previously shown by Mamede et al. (2012), this avalanche size distribution in the dense reservoir network can be described as a power law with exponent ($\alpha$),

$$p_T[X = x] \sim x^{-\alpha}, \tag{4}$$

$$P_T[X \geq x] \sim x^{-\gamma}, \gamma = \alpha - 1, \tag{5}$$

revealing a lack of characteristic avalanche size (see Bak, 1996; Bour and Davy, 1997; Goldstein et al., 2004; Mitzenmacher, 2004; Sornette, 2006). The exponent $\alpha$ of the scale-free avalanche size distribution is estimated using the method of maximum likelihood (MLM) (Clauset et al., 2009) and corresponds to the slope of the probability distribution function. It can be observed in a double logarithmic plot, as shown in the upper right corner of Figure 2. This slope is used to describe the statistics of the clusters of connected reservoirs over a given time $T$. Previous studies (Mamede et al., 2012) focused on describing the scaling exponent of the infinite physical system. In this paper, the emphasis is on the hydrological properties of the real (finite) reservoir network. The generating process of the power law is the competition at the level of each reservoir between water drained into it and water evaporated from its surface. Due to the finite size of the reservoir system, the tail of the distribution shows a clear exponential cutoff.

Based on the avalanche size distribution $p_T[X = x]$ the variable $C_{norm}$ is defined as the connectivity ratio, which corresponds to the maximal number of connected reservoirs within the time window $T$, normalized by the total number of reservoirs

$$C_{norm} = \frac{max_T x}{N_{res}}, \tag{7}$$

where $x$ is the avalanche size, $N_{res}$ is the total number of reservoir, and $T$ is the time interval. High $C_{norm}$ values indicate high vulnerability of the system concerning flood avalanches. From the point of view of water resources, the state of the reservoir system can be described by the variable $V_{norm}$, defined as the maximal amount of water stored in all reservoirs over a given time period $T$, normalized by the total capacity of the network

$$V_{norm} = \frac{max_T V_t}{V_{tot}}, \qquad (8)$$

where $V_t$ is the water volume stored in the reservoir system at time $t \in [0,T]$. In semiarid basins, where temporal variation of stored water is high, low $V_{norm}$ values indicate high vulnerability towards droughts. Both variables, the scaling exponent $\alpha$ and the maximal connectivity $C_{norm}$, are assumed to be strongly dependent on the amount of water in the system $V_{norm}$. Since big clusters of connected reservoirs will arise only after some time of rainy days, the state of the reservoir system is described by the maximum stored volume. For water management purposes it is desirable that the reservoir connectivity ($C_{norm}$) be as low as possible, for stored volume in the system ($V_{norm}$) as high as possible, i.e., the optimum is a state with the lowest possible vulnerability concerning both flood-avalanche and drought hazards. This combination, low $C_{norm}$ and high $V_{norm}$, was a key criterion to assess the performance of the system.

**2.5. Overspill avalanche management**

To understand the network for the purpose of making decisions in the context of water resources management and, more specifically, in the context of the interplay between water availability and flood-avalanche management, additional methods for describing the system behavior were introduced. By identifying single reservoirs that represent special behavior within the network, the reservoir system can be improved in its overall performance. The variable used was

$$N_{b,T} = \frac{N_{buffer,T}}{N_{res}}, \qquad (9)$$

where $N_{buffer,T}$ is the number of reservoirs that are not spilling within the time period $T$ but have at least one spilling reservoir immediately upstream (buffer) and $N_{res}$ is the total number of reservoirs in the system. The expressiveness of $N_{b,T}$ can be augmented by considering the spilling frequencies $F_T$ of the upstream reservoirs that form a connected cluster together with the buffer itself. Frequency of spilling $F_T$ is defined as the fraction of time (within T) a reservoir is spilling. Thus, the importance of the buffers can be quantified and be used for comparison among different topologies. Such a system property was defined as

$$N_{bf,T} = \frac{1}{N_{res}} \cdot \sum_{i=1}^{N_{res}} r_i \cdot F_i^*, \qquad \text{with} \quad F_i^* = \sum_{j \in R_{i,up}} F_{T,j} \qquad (10)$$

where $r_i = 0$ if the reservoir $i$ is spilling, and $r_i = 1$ if not; $F_i^*$ is the sum of the spilling frequencies of all reservoirs directly upstream of reservoir $i$ ($R_{i,up}$).

Thirteen simulation runs were performed, using the reservoir network A with the parameter setup 07 (Table 1) as a reference. The runs differ in the choice of their runoff coefficient $RC$ (between 0.5% and 13%) which was held constant for each run; evaporation and rainfall rates the same as in reference run 07. Thus, the amount of water drained into the reservoirs was varied. This reflects the annual variability in rainfall and/or changes in the runoff coefficient itself. The selected range of $RC$ was based on previous researches in the Brazilian semiarid region. RC changes may be caused, for instance, by changes in climate (Krol et al., 2006; 2011), land use (Gaiser et al., 2003), water management (de Araújo et al., 2004) and/or reservoir network topology (Malveira et al., 2012). From each *RC*, the year with maximum $V_{norm}$ (saturation) has been identified and analyzed. The justification for using this criterion is that the saturation year is the most vulnerable concerning flood avalanches. The system reached saturation at increasing $V_{norm}$ values for increasing RC; whereas the number of simulation years until saturation dropped with increasing RC.

**2.6. Impact of the network on the hydrological connectivity**

2.6.1. Impact of input parameters and network topologies

The manifold uncertainties, in both the reservoir network and the input parameters of the model, are expected to have an impact on the avalanche size distribution. Thus several input parameters were varied to quantify the uncertainty in the avalanche size distribution, specifically seven simulation setups were defined. Every single setup was applied to each of the six network topologies (delineated in 2.3.2), resulting in a total of forty-two simulations (Table 1). Each combination was simulated at daily time steps from 1991 to 2010. Rainfall data for each reservoir and its respective direct catchment area was taken from the closest gauging station (FUNCEME, 2011). The initial runoff coefficient (*RC*) was admitted to be 4% for the whole basin and the same modeling period (Mamede et al., 2012). The remaining parameters were manually calibrated by comparing numerical results with measured data from the monitored reservoirs: correction factors of the runoff coefficient $K_{RC}$ = 1.6 for rainy days and $K_{RC}$ = 0.92 for dry days (see Equation 2); Transmission losses parameter *Kt* = 0.90 if the channel is dry and 0.13 otherwise. The latter calibrated parameter agrees with values measured by Costa et al. (2012) in a 60 km Jaguaribe River reach immediately upstream the Orós reservoir, and also with Hacker (2005).

The input parameter sets have different levels of detail in both the natural variables (rainfall, runoff coefficient, evaporation, and transmission losses) and the anthropogenic/model variables (initial condition, activation time, water release). E.g. Rainfall data refers to measured daily values in setups 01–06 and monthly averaged values in setup 07; In setups 01–03 varying runoff coefficients were considered as well as transmission losses ('trans loss'); The initial conditions were 20% of the capacity (with correction for the monitored reservoirs in setups 01 and 04, and without correction in setups 02 and 05) and empty in setups 03, 06, and 07. In the latter setups, the construction date of the reservoirs has been neglected, and only in setups 01 and 04 the water releases have been considered. The input parameter sets vary from the most detailed (setup 01) over the previously defined (setup 04, Mamede et al., 2012) up to the most generic (setup 07).

2.6.2. Impact of different reservoir classes

As done in other researches (Lima Neto et al., 2011; Krol et al., 2011; Malveira et al., 2012), the role of the small and middle-sized reservoirs in the hydrological system has been investigated. In this research, their impact on flood avalanches was examined. The reservoirs were clustered into five classes (Table 2), as often used (Malveira et al., 2012; Krol et al., 2011; Güntner et al., 2004).For the sake of simplicity, only the reservoir network A with the setup 07 was taken as a reference. Aw1 presents the system A without the reservoirs of class 1, regardless their geographical position in the network. The systems Aw12 and Aw123 are defined in an analogous way, where the index numbers correspond to the classes of reservoirs that have been removed (see Table 1). When a reservoir was removed, direct connections between the reservoirs upstream and the one downstream were promptly (re-)established.

2.6.3. Uniform distribution of capacity and drainage area

Besides the result from varying the input parameters (setups 01-07) and the diverse definitions of the network (topologies A-F, as well as Aw1, Aw12, and Aw123), it is of paramount interest to study the resilience of connectivity patterns to changes in several fundamental properties of the system. The relation between incoming and outgoing water for each reservoir is mainly dependent on its drainage area and the actual water surface because of evaporation losses. Together with eventually spilling upstream reservoirs, the basin size and storage capacity seem to be the most reasonable properties to be investigated. Thus, the capacity and the drainage area of the reservoirs were changed and the resulting system behavior was observed. Network D with parameter set 07 (Table 1) was chosen to be the reference. Three runs were defined: run *orig* (for 'original') denotes the reference. Run *capuni* admitted uniformly distributed reservoir capacity, obtained from total storage capacity (3600 hm³) divided by the number of reservoirs (4237), which yields an average capacity of 0.85 hm³, assigned for each reservoir. The run *basuni* was generated by applying the same methodology with respect to the drainage area, with an outcome of 5.6 km² for each reservoir.

2.6.4. Reallocation of the existing network

To evaluate the sensitivity of the specific distribution of the reservoir capacity in the network, 100 artificial reservoir systems were built. All reservoirs belonging to the same Strahler order of the river network were randomly shuffled and a new capacity was assigned. This is reasonable because the order of the river stream is coupled to the accumulated drainage area (Rodriguez-Iturbe and Rinaldo, 2001) and, therefore, to the amount of water that is expected as inflow. This way, the order of magnitude for the new allocated capacities lies within a feasible range. To bring even more variation into the reservoir networks, every reservoir got then a new location in the system. The new position in the Upper Jaguaribe basin was determined by randomly choosing a new river stream with the same original Strahler order. This way, each reservoir got a new basin and new up- and downstream reservoirs. The third reallocation methodology applied in this study retains the ratio between drainage area and reservoir capacity, but changes the interconnection between reservoirs. The linkage degrees of the reservoirs belonging to the same Strahler stream order were shuffled and new upstream reservoirs were allocated. The simulation used network D and parameter setup 07 (Table 1). These reallocation methods are supposed to reflect generic engineering based rules. The generated random networks still hold important properties of the original reservoir system.

## 3. RESULTS AND DISCUSSION

We characterize the reservoir network and its underlying river network and we quantify the reservoir system with the previously defined system variables. It is shown what part of the reservoir network is responsible for the connectivity pattern observed. In the case of the Upper Jaguaribe basin, the system evolved without integrated planning. To make reasonable decisions in the management of the reservoir system, it is eminent to understand the systems behavior and therefore to identify the reasons for its performance from a hydrological point of view.

## 3.1. River and reservoir networks

The used method to identify the initiation of a natural channel provided results, which corresponded to the river network identified in the hydrography map. The connections between reservoir are determined through the river network, whose fractal nature is related to Horton's laws (Horton, 1945), usually stated in terms of Strahler's ordering system (Rodriguez-Iturbe and Rinaldo, 2001; Tarboton et al., 1988). The results of this investigation showed that the river-network order in the Upper Jaguaribe basin is nine. Let RB be the bifurcation ratio; RL the length ratio (given by the division of the average length of streams of a certain order by the average length of one-order smaller streams); and $D$ the fractal dimension, given by $D = \log RB / \log RL$. For the Upper Jaguaribe basin: RB is 4.14, RL is 2.15, and $D$ is 1.93; showing good agreement with the literature ranges (Tarboton et al., 1988), which are, respectively: 3.3 – 4.7, 1.6 – 2.4, and 1.7 – 2.5. Another indication of the fractal structure of the river network is the fact that the drainage area follows a power law. The scaling exponent ($\alpha = 1.43 \pm 0.02$, see Figure 2) falls into the range of typical values (Rodriguez-Iturbe et al., 1992). The abrupt cutoff is caused by the finite size nature of the drainage system at the Orós reservoir.

Table 2 clearly shows the importance of small reservoirs (classes 1 and 2) with respect to their frequency, which is more than 97% of all impoundments; and the significance of the large reservoirs (classes 4 and 5) with respect to their storage capacity, around 85% of the total network capacity. Like the underlying river network, the reservoir network is represented by a treelike graph without loops, and the ordering method of Strahler can be applied as well. The maximal calculated order is five and the bifurcation ratio is around 5.0. The analysis of the cumulated drainage area of the reservoir network of the Upper Jaguaribe basin has also been established (see Figure 2). The differences between the river and the reservoir networks can be explained by the fact that there is a specific selection of river links to build the network of reservoirs. The increase in the bifurcation ratio (RB) from 4.1 of the river network to 5.0 of the reservoir network reflects this selection. A higher RB means that the number of reservoirs decreases faster than the number of streams for higher orders. Therefore, there are fewer reservoirs having a large cumulative drainage area. One of the most fundamental properties of networks is the degree distribution, which is the distribution of the number of upstream links of the reservoirs. The data of the investigated network shows a power law with scaling exponent of $2.20 \pm 0.05$ (see Figure 2 and Table 1). The first data point shows a small finite-size effect and has not been accounted for the slope estimation. Networks that show a power law in their degree distribution are classified as scale-free networks (Newman, 2005, 2010). The distribution of the reservoir surface also provides a power law (exponent $2.20 \pm 0.01$, see Figure 2 and Table 1). This can be explained because the size of the dam is in accordance with the size of the area that drains directly into the reservoir, the number of upstream reservoirs, and the potential evaporation rate when properly designed. The scaling of the capacity is then a result of the distribution of surface area (see Equation 3).

## 3.2. Impact of the network on the hydrological connectivity

3.2.1. Uncertainty analysis of the scaling exponent

A power-law distribution of flood avalanche sizes reveals that, in most cases, the avalanche evolves a small number of connected reservoirs. However, large scale events also occur, being responsible for the fat tail in the distribution. The found avalanche size slopes of the Upper Jaguaribe reservoir network are within the range 1.95±0.10, consistent with the values found by Mamede et al. (2012; 1.9±0.2, for reservoir system B and parameter setup 04, see Table 1). In fact, the value of the exponent has proven to be robust, provided several minor changes in the network topology and heavy changes in the input parameters. Even under the crude assumption of monthly average rainfall, uniformly distributed over the whole network, the system showed the same global behavior. Therefore, the origin of the avalanche slope must be reasoned only by the underlying network properties themselves. It is worthy though to understand the small differences in estimated scaling exponents for all parameter and network configurations. Each set of exponents, either belonging to the same reservoir system or to the same simulation run, has been tested against each other to verify if they belong to the same sample distribution by applying the Wilcoxon rank-sum test (Maidment, 1993). The only sets where the H0 hypothesis had to be rejected at a significance level of 1% were reservoir network D and setup 07. This deviation is not a numerical effect. In case of network D, most of the very small reservoirs, especially those in the middle of the network – not at the boundary – were disregarded. Thus, it is more likely to find clusters of few connected reservoirs because the small reservoirs connecting the numerous small clusters to a few big ones are widely missing. The result of the special case of simulation setup 07 is to justify as follows: every year, the amount of water flowing into the reservoirs is relatively high, and the evaporation discharges depend on the updated water surface. The amount of drained water brings the network close to the maximal system storage, and the system tends thereafter to its saturation point (after 10 cycles of wet/dry seasons). At this storage level, the transition goes fast from a few small clusters at the beginning of the wet season to large clusters of connected reservoirs in the middle of the wet season, when the system reaches a quasi-permanent regime due to the finite size of the network forming, therefore, only few middle-sized avalanches. Thus, the estimation method (MLM) yields a higher scaling exponent, showing less variability between the different network topologies.

The system behavior has shown considerable dependency on the amount of water in the network ($V_{norm}$), which is mainly regulated by the large reservoirs (Table 2). Most of the small reservoirs get filled within one wet season and dry out during the following dry season. In Figure 3, the setup 07 was applied to the six reservoir-network topologies and the three variables $\alpha$, $V_{norm}$, and $C_{norm}$ are shown; evaluated for $T$ = 1 year. The more water is stored in the system, the more influence has its finite size, resulting in a higher α because of the lower frequency of middle-sized avalanches. The first data point close to 20% falls off that rule owing to initialization processes. The saturation seems to be reached for $V_{norm}$ ~ 0.9. The evolution of $C_{norm}$ shows a monotonous increase with $V_{norm}$. In the lower region, up to $V_{norm}$ ~ 0.6, the increase in $C_{norm}$ flattens between 25% and 30% of full connectivity. After passing $V_{norm}$ ~ 0.6, the maximal connectivity almost doubles, characterizing the first threshold point. In fact, above $V_{norm}$ ~ 0.6 a fast growth, or even a jump in $C_{norm}$ is observable up to 45% connectivity at $V_{norm}$ ~ 0.8. There are a few large reservoirs in the center of the network that spill only after several wet years or after one extremely wet year (Lima Neto et al., 2011). They prevent several middle-sized clusters from being connected. This way, there is a second threshold point in the system, which can be surpassed if there is enough water. The network topologies that were manually corrected (B and C) show a more abrupt and earlier jump in the case of C, and in general a higher connectivity for B (Table 1). Hence, hand-made corrections are considered to be risky since the system behavior might change in an unpredicted way.

3.2.2. Impact of different reservoir classes

Figure 4 shows the results of the impact of different reservoir classes for the Upper Jaguaribe basin. The slopes of the avalanche-size distributions of the several reservoir networks (2.5±1.0) differ from the previous results (1.95 ± 0.10). Above the threshold value of $V_{norm}$ ~ 0.6, the slopes of all networks fall into the known region and are stable until $V_{norm}$ ~ 0.8. This is remarkable, since the differences between A, Aw1, Aw12, and Aw123 are immense, e.g., the network Aw123 consists of only 22 reservoirs, compared to the 4750 of A. The evolution of $C_{norm}$ clearly shows that the networks with smaller reservoirs reach their first threshold point at a higher $V_{norm}$. Also, if there were no small reservoirs, the maximal connected cluster would be obtained after less wet years than with the smaller ones. They are able to restore a relevant amount of water more upstream and to prevent the big reservoirs from spilling and, consequently, triggering much larger avalanches. The analysis of the four reservoir networks (A, Aw1, Aw12, Aw123) shows that the avalanche size distribution, evaluated over the 20 years of simulation, are strongly influenced by the second group of ten years, where $V_{norm}$ > 0.6. Being close to the saturation point, strong finite size effects are observed. This applies for all network topologies; generating a clear data collapse in the avalanche size distribution. This confirms the result of the invariance of the system behavior in sense of connectivity regarding the different reservoir classes, at least above the threshold value of $V_{norm}$ ~ 0.6, which means total storage above 90 mm. These results are comparable with findings from investigations considering sediment redistribution (Lima Neto et al., 2011) or hydrological sustainability (Malveira et al., 2012). The present results show the importance of the small reservoirs, in particular of reservoir class 3 (and less of classes 1 and 2) in retaining large floods. This influence is valued positively, considering the simultaneous capacity of storing a reasonable amount of water (kept upstream) and of preventing flood catastrophes up to a certain degree.

3.2.3. Uniform distribution of capacity and drainage area

The results concerning the behavior of uniformly-distributed capacity and of uniformly-distributed drainage area topologies are presented in Figure 5. It is noticeable that both variables *α* and $C_{norm}$ of run *capuni* (uniform capacity of reservoirs) behave differently from the relatively smooth evolvement observed in other topologies (e.g., see Figures 3, 4 and 5). The reason is that a large amount of upstream reservoirs do not get filled. The thousands of small most upstream reservoirs would be too large (0.85 hm³) for its catchment area, would evaporate excessively and rarely (if ever) spill. Thus, practically no floods would propagate through the system, which is positive considering avalanche prevention. Nonetheless, this behavior would considerably enhance the system vulnerability to water availability in regular years – not to mention during droughts (Mishra and Singh, 2011). However, the connectivity of run *basuni* (uniform direct drainage area of reservoirs) reproduces qualitatively the known dynamics. Nonetheless, the number of maximal connected reservoirs is generally higher than that obtained for the original network; and the jump in $C_{norm}$ (first threshold) occurs at a lower value of $V_{norm}$. The lower slope of the avalanche-size distribution indicates a generally higher possibility of having large connected clusters. The absence of the widely distributed drainage areas flattens the competition between spilling and non-spilling reservoirs, but the primary system behavior does not change. The importance of capacity scaling had also been discussed in the context of the hydrological sustainability of dense-reservoir networks (Malveira et al., 2012). The original (as built) reservoir network performed better than *capuni* or *basuni* runs. This can be stated because the original topology provides lowest $C_{norm}$ (~ 0.45; lowest flood-avalanche vulnerability, thus) for the highest $V_{norm}$ (~ 0.9; lowest water-scarcity vulnerability). The predominant factor for this result is that the capacity distribution of the original network is rationally related with the evaporation from the water surface.

3.2.4. Reallocation of the existing network

In Figure 6, there are the results of 100 artificial reservoir networks composed of the same reservoirs as in A (Table 1), but the capacities were shuffled among all reservoirs belonging to the same Strahler order of the river network. The continuous increase of $\alpha$ with $V_{norm}$ is observable for every single network. The slope is on average slightly higher than in the original network. Besides, the saturation status at $V_{norm}$ ~ 0.9 has not been reached by any synthetic network, showing that the original topology maximizes the effective storage capacity of the system. This general shift of system behavior to lower values of $V_{norm}$ can also be observed in the plot of $C_{norm}$. The fast increase in the maximal connectivity (first threshold) is encountered at lower values of $V_{norm}$ for the artificial than for the original topology. Even more remarkable is that, below the first threshold at $V_{norm}$ ~ 0.6, the original network shows the lowest value of $C_{norm}$, i.e., the lowest vulnerability in terms of flood avalanches. All this information can be put together in the statement that the original reservoir system has naturally evolved into a state where the least possible number of connected reservoirs is found for every value of $V_{norm}$ before a jump in $C_{norm}$ is observable. It is remarkable that the original system has the lowest flood-avalanche vulnerability (lowest $C_{norm}$ ~ 0.45) for the lowest water-scarcity vulnerability (highest $V_{norm}$ ~ 0.9) when compared with any of the 100 artificial topologies. This system performance has been obtained only by applying specific design criteria at the level of single reservoirs, without any systematic strategy of the authorities.

The results of the reservoir networks built by reallocation of each reservoir to a randomly chosen river stream, belonging to the same Strahler order as the original, are presented in Figure 7. The evolution of $\alpha$ and $C_{norm}$ with increasing amount of water captured in the system is similar to the runs, for which the capacities are redistributed. Here, as well, the maximal storage of $V_{norm}$ ~ 0.9 has not been reached by the synthetic networks. This is expected, since the redistribution of the capacity is in fact analogous to the choice of a new location. However, the fast increase in $C_{norm}$ after the first threshold value is either absent or takes place at a very low amount of water in the system. This effect is supposedly lost due to the movement of the large strategically positioned reservoirs to new locations having less relevance for the system behavior. This analysis confirms that the real network builds a lower boundary for the number of maximal connected reservoirs. The statement made before, that the highest approvable performance of the system has been obtained without integrated and systematic consideration, is supported by these additional results. The system behavior for reservoir networks built by random allocation of upstream reservoirs with respect to the linkage degrees among the reservoirs belonging to the same Strahler river stream order is also shown in Figure 7. As observed in previous results, the slope does not change in its evolution with increasing $V_{norm}$. Its average value is slightly above the original one. The maximal number of connected reservoirs of each wet season is always above the real values of $C_{norm}$. In most of the new networks, the jump in $C_{norm}$ can be spotted, sometimes at very low values of $V_{norm}$, sometimes close to $V_{norm}$ ~ 0.6, but not above. Again, the artificial networks do not change considerably the connectivity pattern, but show lower performance than the original one with respect to the maximal number of connected clusters and to the optimized topology (simultaneously lowest vulnerability to flood avalanches and to water scarcity).

3.3. Overspill avalanche management

The results of the 13 runs differing in their constant input runoff coefficient (*RC*) from 0.5% to 13% show that the relation of α (2.10±0.10) is different from the ones in the previous sections because now all data points represent the system only at its saturation. For a decreasing-*RC* scenario, the water stored in the system lowers continuously, but all connectivity properties (including the avalanche-size distribution) proved to be similar whenever a dry year followed a wet year, because no additional upstream reservoir contributed to the downstream reservoir. For the lowest runoff coefficients (*RC* < 2%), revealing $V_{norm}$ values at saturation smaller than 0.4, the system showed larger scaling exponents (α) than those previously observed. Because such low *RC* values are unlikely to be observed in large catchments under the present climatic, hydrologic and land-use constraints, the saturation $V_{norm}$ values could only be caused by droughts and, therefore, flooding is a minor issue in this regime. With more water in the system (*RC* values between 2% and 7%), the slope stabilizes (2.03±0.02) over a broad range of $V_{norm}$, from 0.4 to 0.9. It is within that region that the system presented the jump in the maximal number of connected reservoirs ($C_{norm}$). Above that range, the actual storage would be very close to the maximal possible storage volume and the finite size of the system plays an increasingly important role, leading to a marginal increase of the slope from 2.04 to 2.07.

The measures $N_b$ and $N_{bf}$ are shown for the 13 simulation runs in Figure 8. $N_b$ shows a peak around $V_{norm}$ ~ 0.2 and falls continuously afterwards. Complimentarily, $N_{bf}$ peaks at a much higher water-storage level ($V_{norm}$ ~ 0.6). This means that with more and more water in the network there will be more spilling reservoirs as well and the number of buffers falls, but few remaining buffers compensate the failing of their upstream buffers and can prevent a catastrophic avalanche. At the peak of $N_{bf}$, few buffers fail and the avalanche that had previously been suppressed is triggered. This is the reason for the jump in $C_{norm}$, previously mentioned. Close to the maximum storage capacity of the system, both $N_b$ and $N_{bf}$ fall drastically to zero, when all the reservoirs connect.

A simple strategy to improve the performance of the system concerning the prevention of catastrophic flood avalanches would be to identify the important buffers and to protect them from spilling. The probability that single non-spilling reservoirs in the middle of the network prevent the occurrence of large clusters is higher, the less other upstream connections are available. Hence, the higher the linkage degree, the higher the possibility of a large number of connected reservoirs and the lower the slope of the avalanche size distribution. The absolute values of maximal connected reservoirs only show marginal differences among the systems with varying linkage degree. The distribution of the linkage degree found in the original reservoir system seems to be crucial for an appropriate system performance, i.e., to prevent large avalanches propagating through the system and connecting most of the reservoirs to one large cluster, for example, larger than half the system size. In Figure 9, *F\** (see Equation 10) is displayed for each reservoir over twenty years for setup 01 (Table 1). Because the maximal capacities of the reservoirs were defined based on data collected after heavy rainfalls during three years that are among the 20 simulation years, each reservoir was expected to spill. But not all did. Table 2 summarizes the reservoirs that never spilled for the simulations performed for the network A (see Table 1). This can be seen as a simple check of the model accuracy. In fact, about 10% of all reservoirs never spilled. It is conspicuous that in the most southern and most northern part of the Upper Jaguaribe basin the model predicts numerous non-spilling reservoirs. The big red dots in Figure 9 are large reservoirs that have many frequent spilling reservoirs upstream. They have spilled at least once themselves and eventually connect downstream to big clusters. The system performance would be improved, by means of reducing the probability of a flood avalanche without lowering $V_{norm}$, if these reservoirs could be turned into strategic buffers, e.g., by building upstream reservoirs which act as buffers; by reducing their direct inflow; or by properly operating these reservoirs, especially when control gates would be available (Hsu and Wei, 2007; Li et al., 2010). The latter initiative will not only lower flood-discharge peaks, but also raise evaporation losses before the extreme events, allowing for larger buffer volume. The gray dots in Figure 9 denote non-spilling reservoirs, holding water back and protect the network from even larger avalanches. Thus, one has to be careful when changing their upstream hydrological regime e.g., by destroying small upstream reservoirs because of their low hydrologic efficiency (see Malveira et al., 2012), or by clearing parts of the direct catchment for agricultural purpose, which would lead to an increase of the runoff coefficient.

## 4. CONCLUSIONS

The connectivity pattern during flood events of an existing dense reservoir network in north-eastern Brazil has been investigated. The system, composed of more than 4,000 dams, has been formed without integrated planning of the authorities, except for few large reservoirs. By spilling of single reservoirs, parts of the system get connected and flood avalanches propagate throughout the system. Even with the immense variety of approaches considered here to rebuild the system artificially, the slope in the avalanche-size distribution, which defines the scaling of the number of connected reservoirs, is found to be in the range 1.95±0.10. This range has its origin in long term parameters, as the reservoir network itself, its drainage network (the reservoir network can be seen as an overlay of the river network) and the corresponding landscape, which has been formed over millions of years. By generating artificial network topologies within certain constraints (given by the original network), the genuine topology performed best with respect to the maximal observed flood-avalanche sizes. The most sensitive properties – concerning the system vulnerability to flood avalanches – are the linkage degree distribution and the presence of the small and middle-sized reservoirs. The found linkage degree distribution suppresses large avalanches and connects most of the reservoirs to big clusters, which can be larger than half of the system size. Our results are comparable with findings from investigations considering sediment redistribution (Lima Neto et al., 2011) and hydrological sustainability (Malveira et al., 2012); and show the importance of the small reservoirs, in particular of reservoirs from class 3 (1 to 10 hm³) in retaining large floods. The original reservoir network has evolved into a state that could not be reproduced by artificial networks built by reasonable redistribution of several reservoir parameters. Surprisingly, the optimal configuration of the system has been reached in real world, only by applying design rules for single reservoirs, without an integrated view of the network. The system has, therefore, evolved into this optimal configuration without external fine tuning. This characteristic is a fingerprint of self-organized criticality (Bak, 1996; Sornette, 2006), which has previously been identified in the dense reservoir network by Mamede et al. (2012).The statement made before, that the highest approvable performance of the system was reached without integrated and systematic consideration is fully supported by the additional results of this research.

Future work might apply the concepts and methods introduced to other existing dense reservoir networks. In addition, it is interesting to analyze the effect of time and spatial correlations of the rainfall on the overall dynamics. The inter-seasonal variations and the impact of specific chronologies of dry and wet years are also of interest. To make decisions at the scale of single reservoirs, the spatial variability of main hydrological parameters has to be included, e.g. spatial varying runoff coefficient representing soil properties and vegetation cover.

**ACKNOWLEDGEMENTS**

The authors would like to acknowledge the Swiss Federal Institute of Technology Zurich - ETH and the Brazilian National Council of Scientific and Technologic Research - CNPq (200672/2012-6), for the support to this research. The authors would also like to thank Prof. George L. Mamede for the valuable discussions.

**REFERENCES**

Andrade, E.M., Palácio, H.A., Souza, I.H., Leão, R.A., Guerreiro, M.J., 2008. Land use effects in groundwater composition of an alluvial aquifer (Trussu River, Brazil) by multivariate techniques. Environ. Res. 106, 170-177

Bak, P., 1996. How nature works, the science of self-organized criticality. Copernicus, Springer-Verlag, New-York.

Boroto, R.A.J., Gôrgens, A.H.M., 2003. Estimating transmission losses along the Limpopo River: an overview of alternative methods. IAHS Publ. 278, 138 – 143

Bour, O., Davy, P. 1997. Connectivity of random fault networks following a power law fault length distribution. Water Resour. Res. 33(7), 1567-1583

Campos, J.N.B., 2010. Modeling the yield-evaporation-spill in the reservoir storage process: the Regulation Triangle Diagram. Water Resour. Manage. 24, 3487-3511

Clauset, A., Shalizi, C.R., Newman, M.E.J., 2009. Power-law distributions in empirical data. SIAM Rev. 51(4), 661–703

COGERH - Companhia de Gestão dos Recursos Hídricos. http://www.cogerh.com.br, accessed 2011.

Costa, A.C., Bronstert, A., de Araújo, J.C., 2012. A channel transmission losses model for different dryland rivers. Hydrol. and Earth Syst. Sci. 16, 1111-1135

de Araújo, J.C., Lima Neto, I.E., Medeiros, P.H.A., Malveira, V.T.C., 2010. Impact of a dense reservoir network on water availability in the semiarid north-eastern Brazil, IAHS Hydropredict 2010, Prag. http://iahs.info/conferences/CR2010/2010_Praha/full/024.pdf

de Araújo, J.C., Piedra, J.I.G., 2009. Comparative hydrology: analysis of a semiarid and humid tropical watershed. Hydrol. Process., 23, 1169-1178, doi:10.1002/hyp.7232

de Araújo, J.C., Güntner, A., Bronstert, A., 2006. Loss of reservoir volume by sediment deposition and its impact on water availability in semiarid Brazil. Hydrol. Sci. J. – J. Sci. Hydrol. 51(1), 157–170

de Araújo, J.C., Döll, P., Güntner, A., Krol, M., Abreu, C.B.R., Hauschild, M., Mendiondo, E.M., 2004. Water scarcity under scenarios for global climate change and regional development in semiarid Northeastern Brazil. Water Int. 29(2), 209-220

Dutta, D., Herathb, S., Musiake, K., 2003. A mathematical model for flood loss estimation. J. Hydrol. 277, 24-49

FUNCEME - Fundação Cearense de Meteorologia e Recursos Hídricos. http://www.funceme.br, accessed 2011

Gaiser, T., Krol, M.S., Frischkorn, H., de Araújo, J.C., 2003. Global change and regional impact. 1st ed. Springer Verlag, Berlin

García-Pintado, J., Barberá, G.G., Erena, M., Castillo, V.M., 2009. Calibration of structure in a distributed forecasting model for a semiarid flash flood: Dynamic surface storage and channel roughness. J. Hydrol. 377, 165–184

Goldstein, M.L., Morris, S.A., Yen, G.G., 2004. Problems with fitting to the power-law distribution. Eur. J. of Phys. B 41(2), 255–258

Güntner, A., Bronstert, A., 2004. Representation of landscape variability and lateral redistribution processes for large-scale hydrological modelling in semi-arid areas. J. Hydrol. 297(1–4), 136-161

Güntner, A., Krol, M.S., de Araújo, J.C., Bronstert, A., 2004. Simple water balance modelling of surface reservoir systems in a large data-scarce semiarid region. Hydrol. Sci. J. – J. Sci. Hydrol. 49(5), 901–918

Gundekar, H.G., Khodke, U.M., Sarkar, S., Rai, R.K., 2008. Evaluation of pan coefficient for reference crop evapotranspiration for semi-arid region. Irrig. Sci. 26(2), 169–175

Hacker, F., 2005. Model for water availability in semi-arid environments (WASA): estimation of transmission losses by infiltration at rivers in the semi-arid Federal State of Ceará (Brazil). Master's thesis, Universität Potsdam.

Horton, R., 1945. Erosional development of streams and their drainage basins; hydrophysical approach to qunatitative morphology. Geol. Soc. of Am. Bull. 56(3), 275–370

Hsu, N.-S., Wei, C.-C., 2007. A multipurpose reservoir real-time operation model for flood control during typhoon invasion. J. Hydrol. 336, 282– 293

Ijjasz-Vasquez, E., Bras, R.L., 1995. Scaling regimes of local slope versus contributing area in digital elevation models. Geomorphol. 12(4), 299–311

Jenson, S.K., Domingue, J.O., 1998. Extracting topographic structure from digital elevation data for geographic information system analysis. Photogramm. Eng. and Remote Sens. 54(11), 1593–1600

Krol, M.S., Jaeger, A., Bronstert, A., Güntner, A., 2006. Integrated modelling of climate, water, soil, agricultural and socio-economic processes: A general introduction of the methodology and some exemplary results from the semi-arid north-east of Brazil. J. Hydrol. 328(3–4), 417-431

Krol, M.S., de Vries, M.J., van Oel, P.R., de Araújo, J.C., 2011. Sustainability of small reservoirs and large scale water availability under current conditions and climate change. Water Resour. Manage. 25, 3017-3026

Li, X., Guo, S., Liu, P., Chen, G., 2010. Dynamic control of flood limited water level for reservoir operation by considering inflow uncertainty. J. Hydrol. 391, 124–132

Lima Neto, I.E., Wiegand, M.C., de Araújo, J.C., 2011. Sediment redistribution due to a dense reservoir network in a large semi-arid Brazilian basin. Hydrol. Sci. J. – J. Sci. Hydrol. 56, 319-333

Lowe, L., Nathan, R.J., Morden, R., 2005. Assessing the impact of farm dams on streamflows, Part II: regional characterisation. Australian Journal of Water Resources, 9(1), 13-26

Maidment, D.R., 1993. Handbook of Hydrology, McGraw-Hill, Inc., New York.

Malveira, V.T.C., 2009. Pequena açudagem e sustentabilidade hidrológica em grandes bacias semi-áridas: estudo de caso da bacia do açude Orós. PhD Thesis, Universidade Federal do Ceará.

Malveira, V.T.C., de Araújo, J.C., Güntner, A., 2012. Hydrological impact of a high-density reservoir network in the semiarid north-eastern Brazil. J. Hydrol. Eng. 17, 109-117

Mamede, G.L., Araújo, N.A.M., Schneider, C.M., de Araújo, J.C., Herrmann, H.J., 2012. Overspill avalanching in a dense reservoir network. Proc. of the Natl. Acad. of Sci. of the USA 109, 7191-7195

Michaelides, K., Chappell, A., 2009. Connectivity as a concept for characterizing hydrological behavior. Hydrol. Process. 23, 517-522

Mishra, A.K., Singh, V.P., 2011. Drought modeling – A review. J. of Hydrol. 403(1–2), 157–175

Mitzenmacher, M., 2004. A brief history of generative models for power law and lognormal distributions. Internet Math. 1(2), 226–251

Molle, F., 1989. Evaporation and infiltration losses in small dams (in Portuguese), SUDENE, Recife.

Molle, F., Cadier, E., 1992. Manual of the small dam (in Portuguese). ORSTOM-SUDENE, Recife.

Montenegro, S., Ragab, R., 2012. Impact of possible climate and land use changes in the semi-arid regions: a case study from north eastern Brazil. J. Hydrol. 434–435, 55-68. Doi 10.1016/j.jhydrol.2012.02.036

Nathan, R.J., Jordan, P., Morden, R., 2005. Assessing the impact of farm dams on streamflows, Part I: development of simulation tools. Australian Journal of Water Resources 9(1), 1-12

Newman, M.E.J., 2010. Networks - An Introduction. Oxford University Press, Oxford.

Newman, M.E.J., 2005. Power laws, pareto distributions and zipf's law. Contemp. Phys. 46(5), 323–351

O'Callaghan, J.F., Mark, D.M., 1984. The extraction of drainage network from digital elevation data. Comput. Vis., Graph., and Image Process. 28(3), 232–344

Pilgrim, D.H., Chapman, T.G., Doran, D.G., 1988. Problems of rainfall-runoff modelling in arid and semiarid regions. Hydrol. Sci. J. – J. Sci. Hydrol. 33(4), 379 – 400

Rodriguez-Iturbe, I., Ijjasz-Vasquez, E.J., Bras, R.L., Tarboton, D.G., 1992. Power law distributions of discharge mass and energy in river basins. Water Resour. Res. 28(4), 1089–1093

Rodriguez-Iturbe, I., Rinaldo, A., 2001. Fractal river basins, chance and self-organization. Cambridge University Press, Cambridge.

Sornette, D., 2006. Critical phenomena in natural sciences. Chaos, fractals, self-organization and disorder: Concepts and tools, second ed. Springer Verlag, Berlin.

SRTM - Shuttle Radar Topography Mission. http://www2.jpl.nasa.gov/srtm/, accessed 2011.

Tarboton, D.G. 1997. A new method for the determination of flow directions and upslope areas in grid digital elevation models. Water Resour. Res. 33(2), 309–319

Tarboton, D.G., Bras, R.L., Rodriguez-Iturbe, I., 1991. On the extraction of channel networks from digital elvation data. Hydrol. Process. 5(1), 81–100

Tarboton, D.G., Bras, R.L., Rodriguez-Iturbe, I., 1988. The fractal nature of river networks. Water Resour. Res. 24(8), 1317–1322

Zhou, Q., Mikkelsen, P.S., Halsnæs, K., Arnbjerg-Nielsen, K., 2012, Framework for economic pluvial flood risk assessment considering climate change effects and adaptation benefits. J. Hydrol. 414–415, 539–549

**TABLES**

Table 1. Network topologies and variable setups used in the analysis. In the upper part of the table the network topologies A – F are summarized: Nres is the number of reservoirs; RB is the bifurcation rate; Scaling exponents of the power laws are shown for cumulative drainage area, linkage degree, maximal reservoir water surface, and reservoir capacity size. In the lower part of the table the variable setups are presented.

| Network ID | Nres | Total storage capacity (hm³) | (mm) | Average scaling exponents (α) drain area | link | surface | capacity | RB |
|---|---|---|---|---|---|---|---|---|
| A | 4750 | 3,618 | 152.5 | 1.75 | 2.17 | 2.22 | 1.78 | 4.8 |
| B | 3978 | 3,532 | 142.8 | 1.83 | 2.21 | 2.19 | 1.76 | 5.2 |
| C | 3971 | 3,532 | 148.8 | 1.78 | 2.18 | 2.19 | 1.76 | 4.9 |
| D | 4237 | 3,606 | 151.9 | 1.94 | 2.29 | 2.21 | 1.77 | 8.5 |
| E | 4359 | 3,611 | 152.1 | 1.87 | 2.24 | 2.21 | 1.78 | 6.7 |
| F | 4536 | 3,612 | 152.2 | 1.74 | 2.18 | 2.21 | 1.78 | 5.1 |
| Aw1 | 756 | 3,538 | 149.1 | 1.81 | 2.02 | 2.22 | 1.78 | 6.3 |
| Aw12 | 129 | 3,354 | 141.3 | 1.82 | 1.86 | 1.96 | 1.69 | 5.4 |
| Aw123 | 22 | 3,060 | 128.9 | 1.85 | - | 1.86 | 1.86 | 6.3 |

Variables setup

| Setup ID | Natural variables rainfall | RC | evaporation | trans loss | Anthropogenic / modeling variables initial condition | res activ | wat release |
|---|---|---|---|---|---|---|---|
| 01 | stations | Eq. 2 | monthly | yes | 0.20/correction | original | yes |
| 02 | stations | Eq. 2 | monthly | yes | 0.20 | original | no |
| 03 | stations | Eq. 2 | monthly | yes | empty | all | no |
| 04 | stations | constant 4% | monthly | no | 0.20/correction | original | yes |
| 05 | stations | constant 4% | monthly | no | 0.20 | original | no |
| 06 | stations | constant 4% | monthly | no | empty | all | no |
| 07 | monthly | constant 4% | monthly | no | empty | all | no |

Table 2. Classification of identified reservoirs in the Upper Jaguaribe basin with respect to their capacity size for network A (see Table 1). The frequency and the total capacity of the reservoir classes are presented, as well as the number of reservoirs that never spilled during the simulation runs for the period 1991 – 2010.

| Reservoir class | Storage capacity (hm³) | Class frequency (-) | (%) | Class capacity (hm³) | (%) | Number of reservoirs that never spilled |
|---|---|---|---|---|---|---|
| 1 | < 0.1 | 3994 | 84.1 | 80 | 2.2 | 268 |
| 2 | 0.1 – 1 | 627 | 13.2 | 184 | 5.1 | 167 |
| 3 | 1 – 10 | 107 | 2.2 | 294 | 8.1 | 32 |
| 4 | 10 – 100 | 19 | 0.4 | 622 | 17.2 | 8 |
| 5 | > 100 | 3 | 0.1 | 2438 | 67.4 | 1 |
| Total | (-) | 4750 | 100.0 | 3618 | 100.0 | 476 |

**FIGURES**

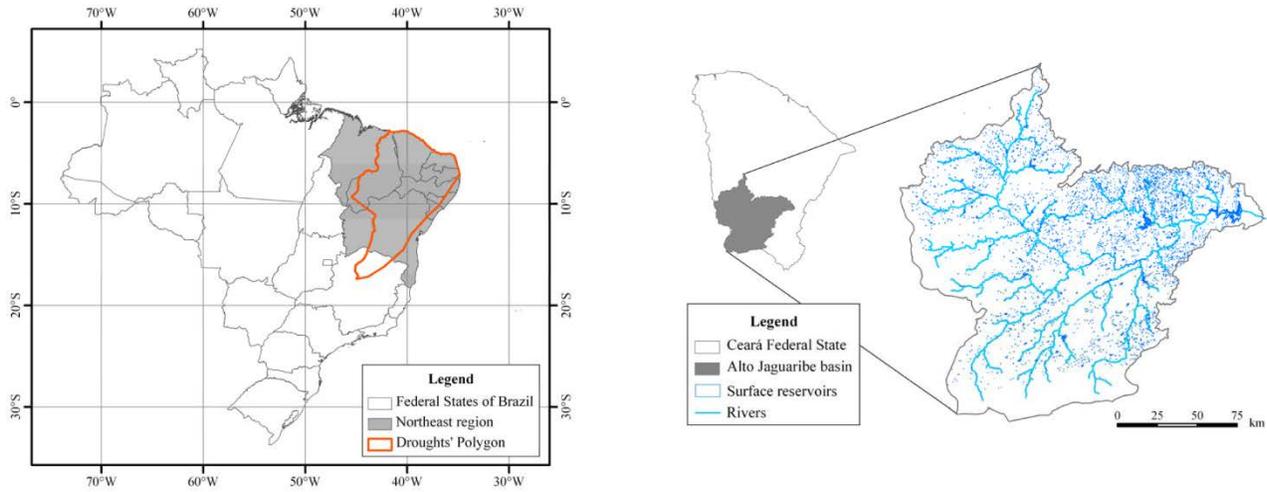

**Figure 1.** Overview of the studied area: (a) The Brazilian semiarid drought polygon, and (b) Upper Jaguaribe Basin, with main rivers and surface reservoirs (Malveira, 2009).

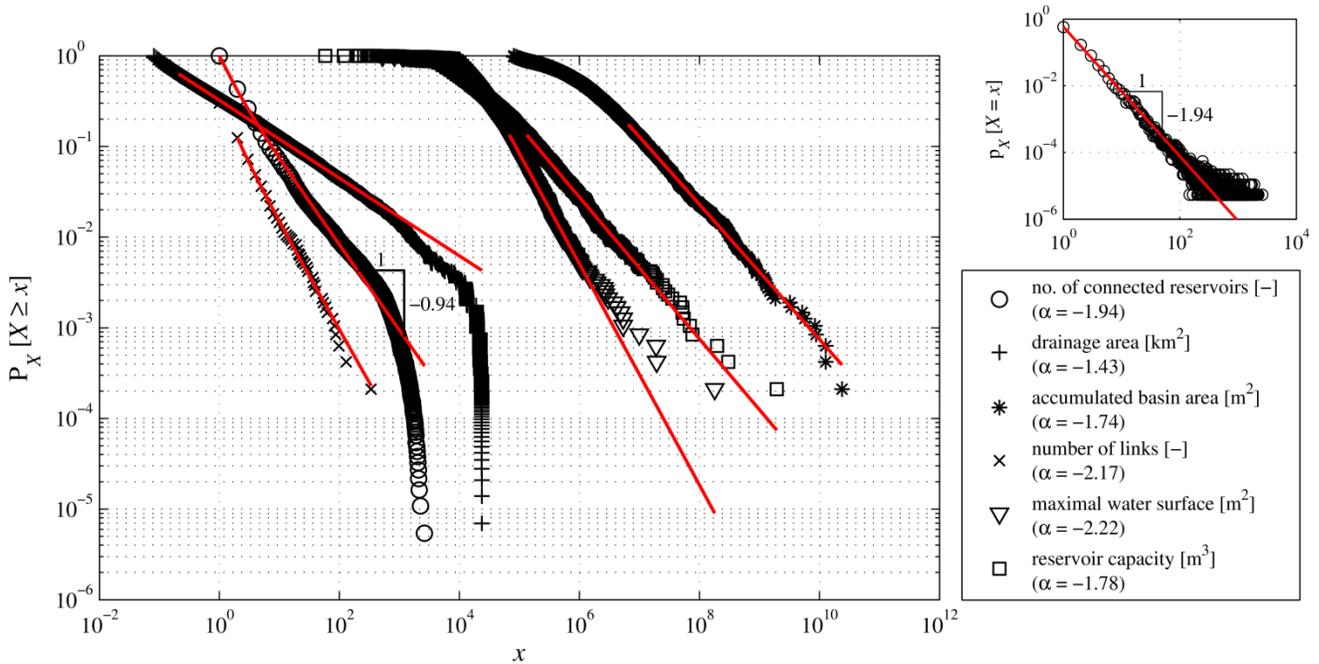

**Figure 2.** The cumulative frequency $P_X[X \geq \chi]$ plot for six reservoir-system variables ($\chi$) and their respective scaling exponents ($\alpha$), applied to network topology A. In the small box (top right) the non-cumulative frequency $p_X[X = \chi]$ plot of the number of connected reservoirs is also presented. Exponent $\alpha$ corresponds to the slope of the distribution measured in a double logarithmic plot.

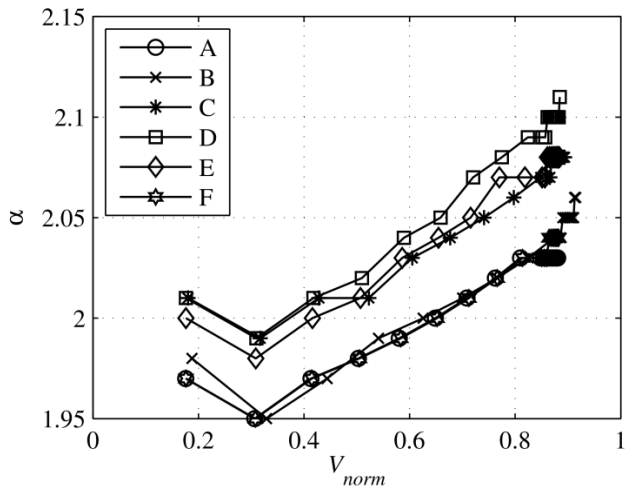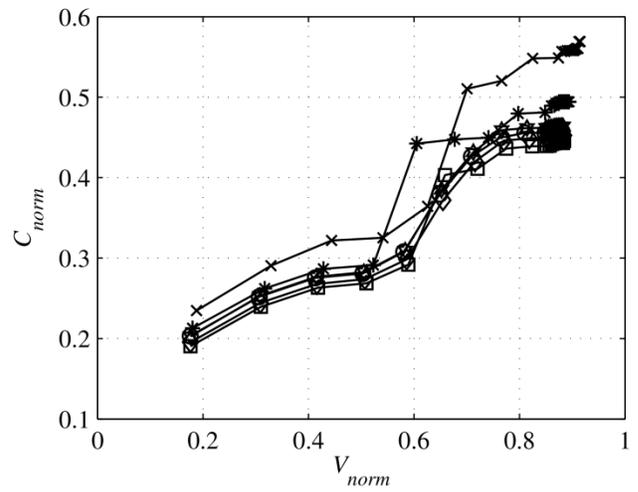

**Figure 3.** Comparison between the actual amount of water in the system ($V_{norm}$) and the two variables $\alpha$ and $C_{norm}$. Each evaluation point represents one wet/dry cycle (1 year).

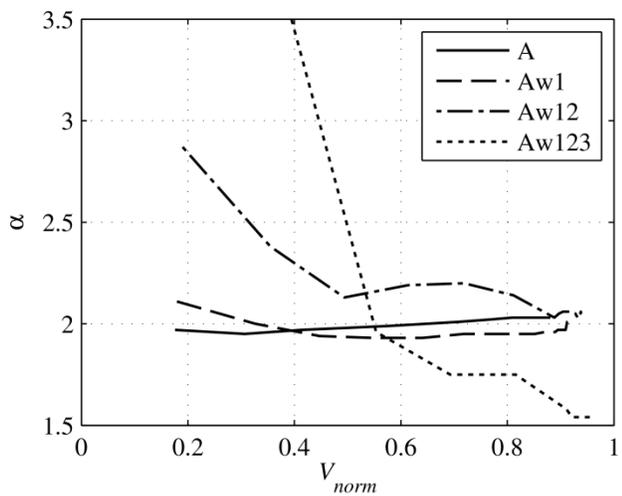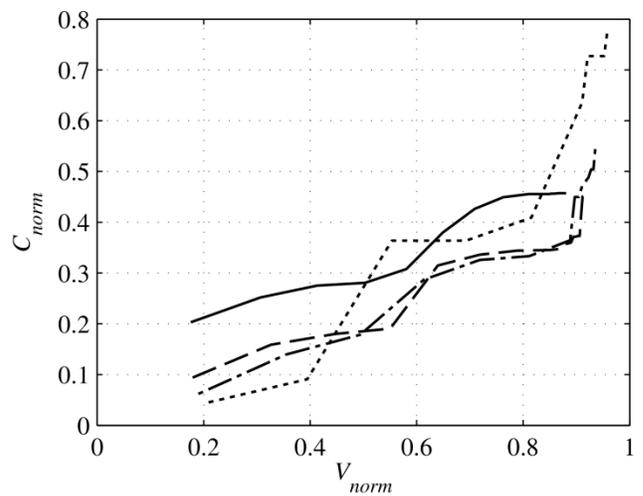

**Figure 4.** Simulation runs with reservoir topologies in which all reservoirs of class 1 (Aw1), all of classes 1 and 2 (Aw12), and all of classes 1, 2, and 3 (Aw123) are neglected: $\alpha$ and $C_{norm}$ are evaluated for each year and plotted against $V_{norm}$.

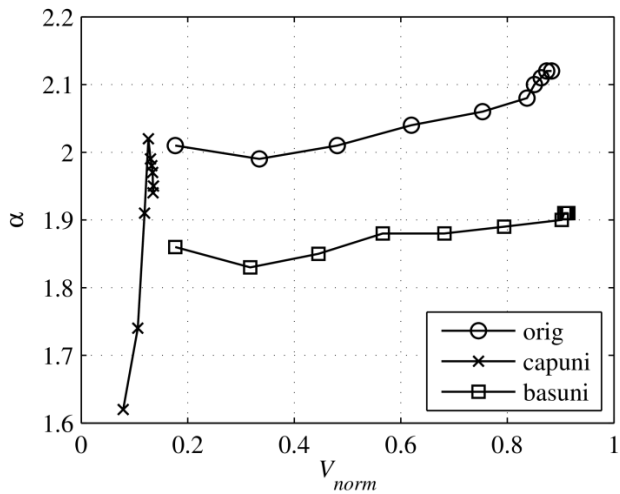 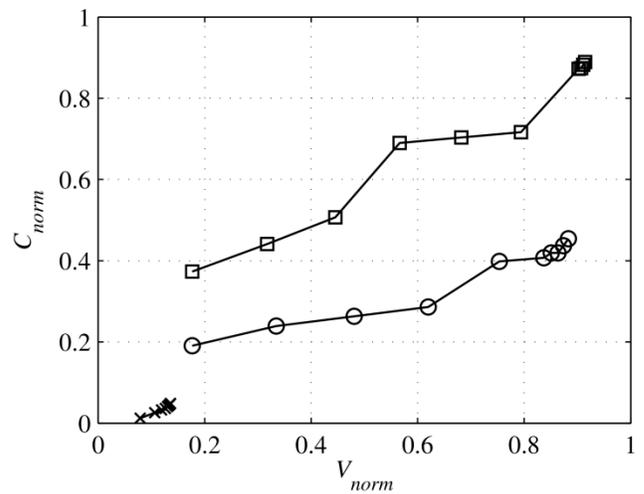

**Figure 5.** Behavior of variables $\alpha$ and $C_{norm}$, taken after each wet/dry season, for network topologies original (*orig*), uniformly-distributed reservoir capacity (*capuni*), and uniformly-distributed drainage area (*basuni*).

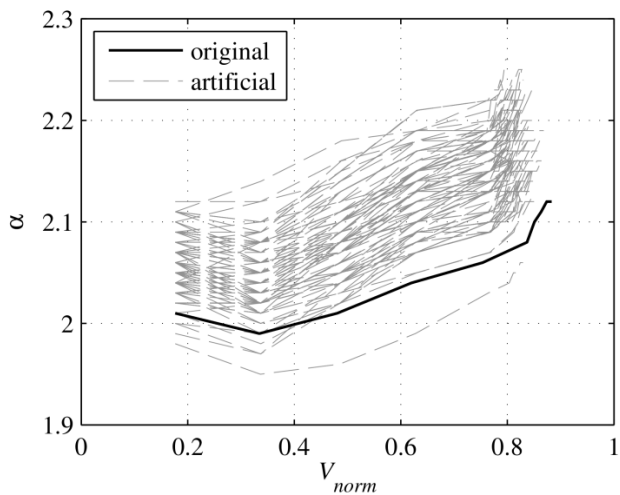 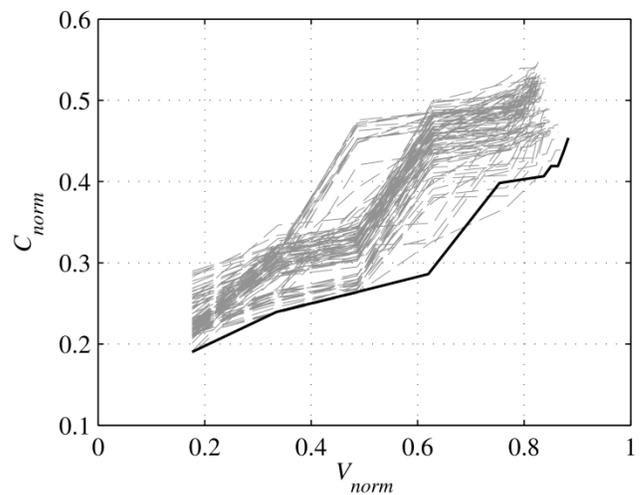

**Figure 6.** Behavior of variables $\alpha$ and $C_{norm}$ for 100 artificial redistributions of reservoir capacities among the reservoirs having the same Strahler order of the assigned river stream (gray lines); in comparison with the original topology (black line).

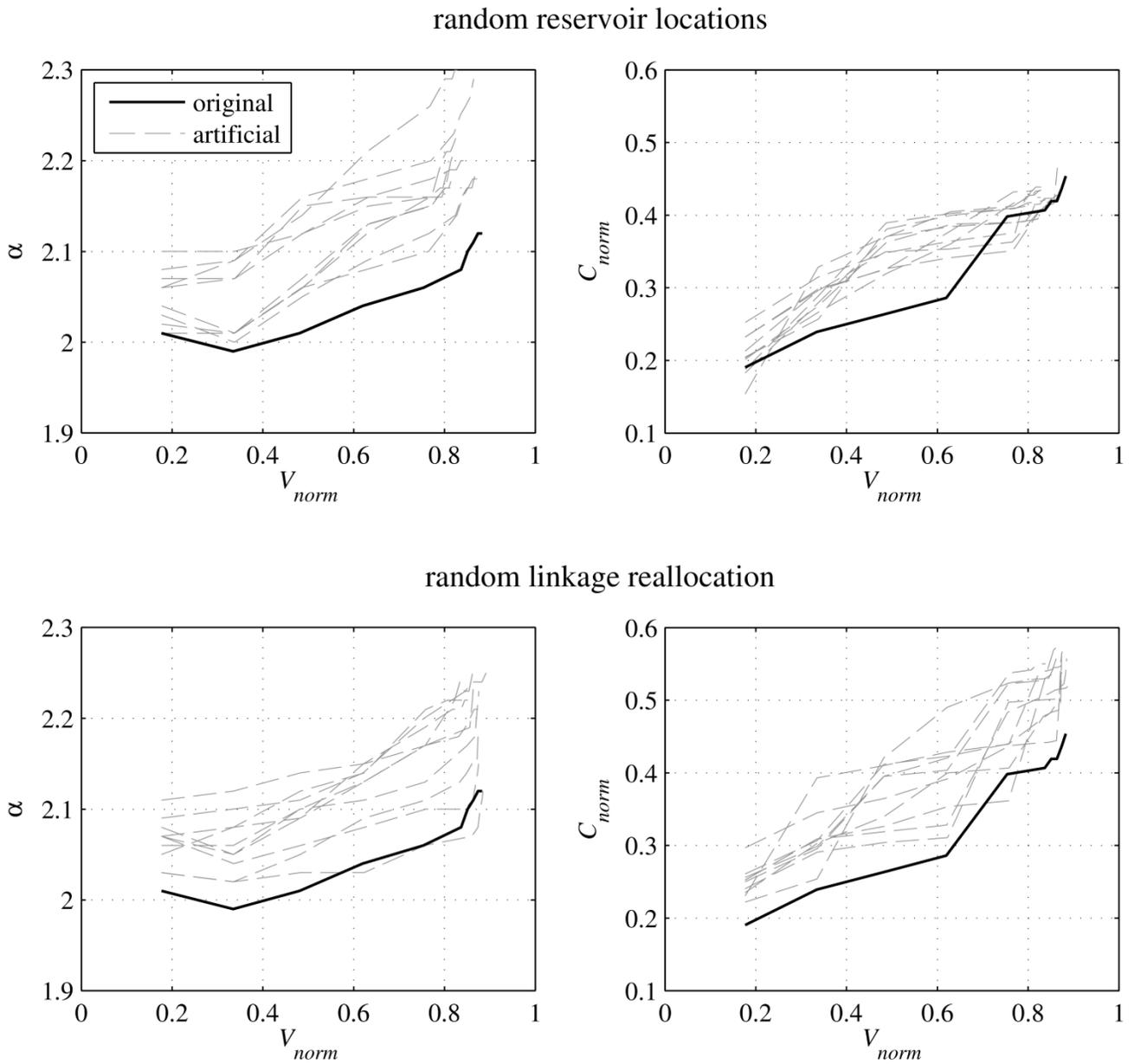

**Figure 7.** Behavior of variables $\alpha$ and $C_{norm}$ for ten topologies built by reallocation of each reservoir to a randomly chosen river stream of the same Strahler order as the original (gray lines, top graphics); for ten topologies built by random allocation of upstream reservoirs with respect to the linkage degrees among the reservoirs belonging to the same Strahler river stream order (gray lines, lower graphics); and for the original run (black lines).

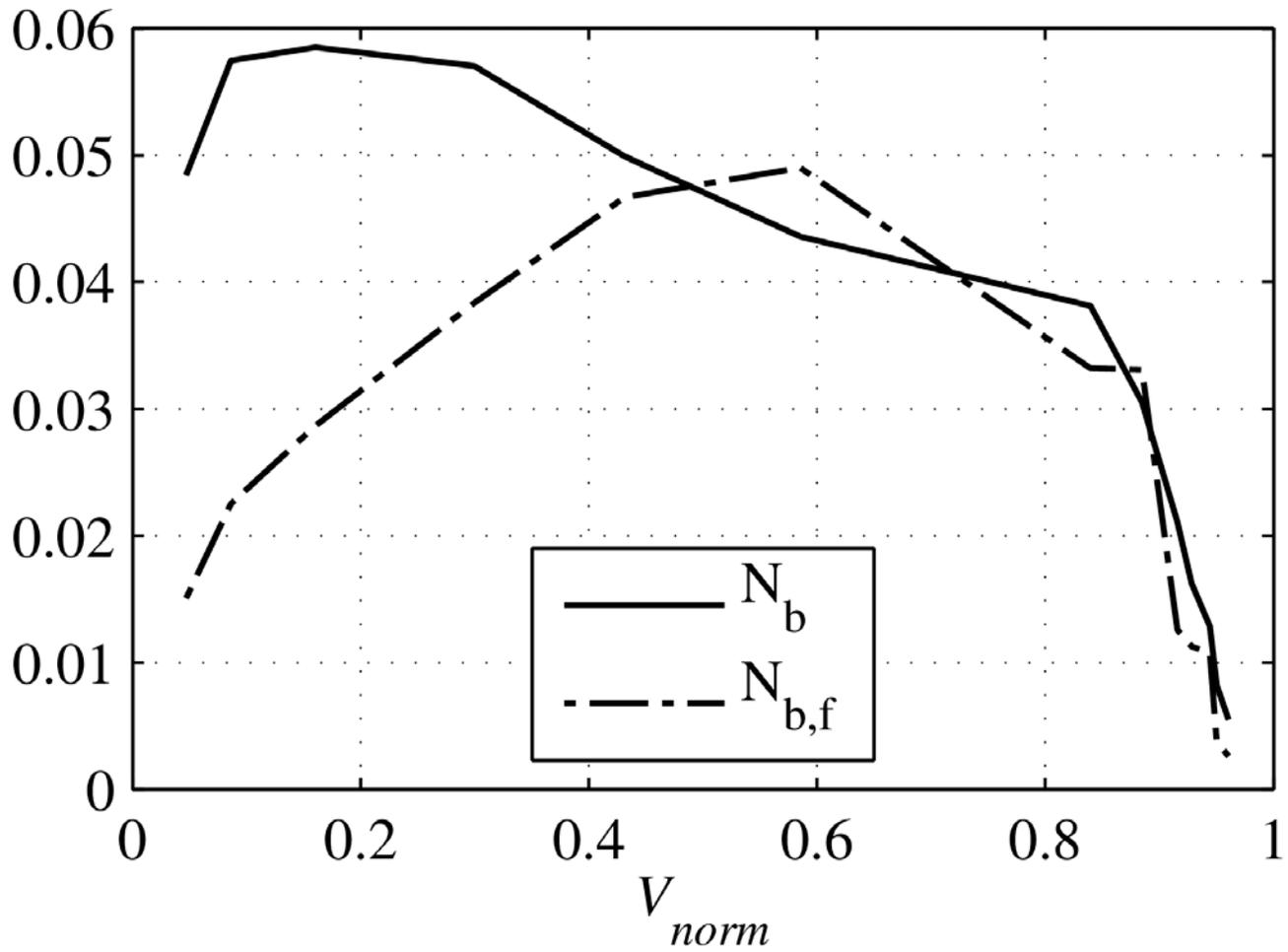

**Figure 8.** Comparative behavior of non-dimensional buffer variables $N_b$ and $N_{bf}$ for the 13 simulation runs with *RC* varying from 0.5% to 13% for each run. Nothe that *RC* was kept constant in space and time throughout each simulation run.

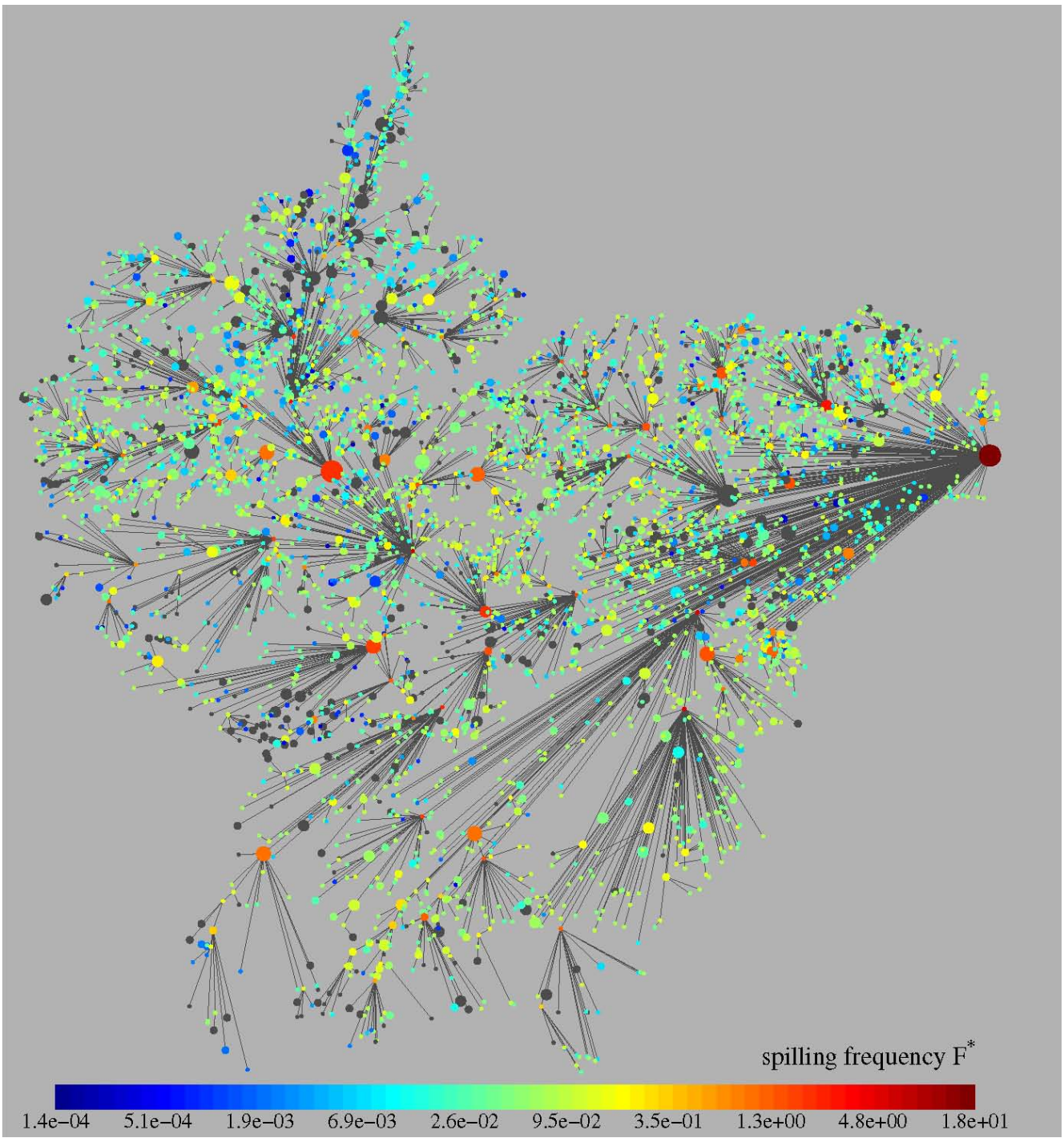

**Figure 9.** Behavior of spilling frequency for network topology A and simulation setup 01: all reservoirs that spilled at least once during the simulation period (1991-2010) are colored with respect to their spilling frequency F* (see Equation 10). The gray reservoirs never spilled.